\PassOptionsToPackage{table}{xcolor}
\documentclass[sigconf]{acmart}
\usepackage[english]{babel}
\usepackage{xspace}
\usepackage{booktabs}
\usepackage{subfigure}
\usepackage{todonotes} %
\usepackage{amsmath}
\usepackage{pifont}
\usepackage{booktabs}
\usepackage{color}
\usepackage{graphicx} %
\usepackage{url}
\usepackage{siunitx}

\usepackage{breakurl}
\usepackage{multirow}
\usepackage[table]{xcolor}
\usepackage{array}  %
\usepackage{soul}
\usepackage{pgf}
\usepackage{xparse}
\definecolor{lightblue}{rgb}{0.68, 0.85, 0.9}
\definecolor{lightred}{rgb}{1.0, 0.6, 0.6}
\definecolor{lightgreen}{rgb}{0.0, 0.8, 0.6}

\copyrightyear{2024}
\acmYear{2024}
\setcopyright{rightsretained}
\acmConference[IMC '24]{Proceedings of the 2024 ACM Internet Measurement Conference}{November 4--6, 2024}{Madrid, Spain}
\acmBooktitle{Proceedings of the 2024 ACM Internet Measurement Conference (IMC '24), November 4--6, 2024, Madrid, Spain}\acmDOI{10.1145/3646547.3689021}
\acmISBN{979-8-4007-0592-2/24/11}

\makeatletter
\gdef\@copyrightpermission{
   \begin{minipage}{0.3\columnwidth}
     \href{https://creativecommons.org/licenses/by-nc-sa/4.0/}{\includegraphics[width=0.90\textwidth]{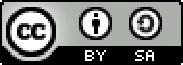}}
   \end{minipage}\hfill
   \begin{minipage}{0.7\columnwidth}
     \href{https://creativecommons.org/licenses/by-nc-sa/4.0/}{This work is licensed under a Creative Commons Attribution-ShareAlike International 4.0 License.}
   \end{minipage}
   \vspace{5pt}
}
\makeatother

\providecommand{\ie}{\emph{i.e.,} }
\providecommand{\eg}{\emph{e.g.,} }

\providecommand{\etal}{\textit{et al}. }

\providecommand{\etc}{\emph{etc.}}   

\newcommand{\dns}[1]{{\small \texttt{#1}}}

\ExplSyntaxOn
\NewDocumentCommand{\abbreviateNumber}{m}
{
    \int_compare:nTF { #1 < 1000 }
    {
        #1
    }
    {
        \int_compare:nTF { #1 < 1000000 }
        {
            \fp_eval:n { round(#1/1000, 1) }~K
        }
        {
            \int_compare:nTF { #1 < 1000000000 }
            {
                \fp_eval:n { round(#1/1000000, 1) }~M
            }
            {
                \fp_eval:n { round(#1/1000000000, 1) }~B
            }
        }
    }
}
\ExplSyntaxOff

\ExplSyntaxOn
\NewDocumentCommand{\abbreviateNumberToThousands}{m}
{
            \int_compare:nTF { #1 < 1000 }
            {
                #1
            }
            {
                \fp_eval:n { round(#1/1000, 1) }~K
            }
}
\ExplSyntaxOff

\begin{document}
\title[DarkDNS: Revisiting the Value of Rapid Zone Updates]{DarkDNS: Revisiting the Value of Rapid Zone Update}

\author{Raffaele Sommese}
\affiliation{
   \institution{University of Twente}
   \city{Enschede}
   \country{The Netherlands}
}
\email{r.sommese@utwente.nl}

\author{Gautam Akiwate} 
\affiliation{
   \institution{Stanford University}
   \city{Stanford}
   \state{CA}
   \country{USA}
}
\email{gakiwate@cs.stanford.edu}

\author{Antonia Affinito}
\affiliation{
   \institution{University of Twente}
   \city{Enschede}
   \country{The Netherlands}
}
\email{a.affinito@utwente.nl}

\author{Moritz M\"uller}
\affiliation{
   \institution{SIDN Labs}
   \city{Arnhem}
   \country{The Netherlands}
}
\affiliation{
   \institution{University of Twente}
   \city{Enschede}
   \country{The Netherlands}
}
\email{moritz.muller@sidn.nl}

\author{Mattijs Jonker}
\affiliation{
   \institution{University of Twente}
   \city{Enschede}
   \country{The Netherlands}
}
\email{m.jonker@utwente.nl}

\author{KC Claffy}
\affiliation{
   \institution{CAIDA}
   \city{San Diego}
   \state{CA}
   \country{USA}
}
\email{kc@caida.org}

\renewcommand{\shortauthors}{Raffaele Sommese et al.}



\begin{abstract}
    Malicious actors exploit the DNS namespace to launch spam campaigns, phishing
attacks, malware, and other harmful activities.  Combating these threats
requires visibility into domain existence, ownership and nameservice 
activity that the DNS protocol does not itself provide.
To facilitate visibility and security-related study of the expanding gTLD
namespace, ICANN introduced the Centralized Zone Data Service (CZDS) that shares
daily zone file snapshots of new gTLD zones.  However, a remarkably high
concentration of malicious activity is associated with domains that do not live
long enough make it into these daily snapshots.
Using public and private sources of newly observed domains, we discover that 
even with the best available data there is a considerable visibility gap in 
detecting short-lived domains.
We find that the daily snapshots miss at least 1\% of newly registered and
short-lived domains, which are frequently registered with likely malicious
intent.
In reducing this critical visibility gap using public sources of data, we
demonstrate how more timely access to TLD zone changes can provide valuable
data to better prevent abuse.
We hope that this work sparks a discussion in
the community on how to effectively and safely revive the concept of sharing
Rapid Zone Updates for security research.
Finally, we release a public live feed of newly registered domains, 
with the aim of enabling further research in abuse identification.

\end{abstract}

\maketitle

\section{Introduction}
\label{sec:intro}

Malicious actors exploit (abuse) the DNS namespace to launch spam campaigns,
phishing attacks, malware, and other harmful activities.  In many dimensions the
DNS ecosystem is more opaque than other aspects of Internet transport. Unlike
BGP, DNS is a pull protocol, so learning internal dynamics requires an entry
point \ie a domain name. 
Without knowing this entry point, any abuse that lies
behind a domain remains opaque to everyone except the targets.

This inherent opacity has motivated many attempts to improve
visibility into the DNS ecosystem to facilitate research,
analysis, and operational security.  A prominent
example is the ICANN Centralized Zone Data Service (CZDS) which
coordinates collection and restricted sharing of daily zone
file snapshots of participating TLDs. 
However, previous studies 
have found that the 24-hour zone 
files, while an unprecedented enabler of DNS-related security research,
leave vital visibility gaps in studies of abuse and exploits
\cite{2020-akiwate-ui,2021-akiwate-rb, ssac-rnm-2024}.
Since so much malicious activity often begins soon after a domain is registered
for that purpose, and the CZDS snapshots are updated only daily, commercial
threat intelligence services~\cite{dt-sie} have tried to fill the gap by
detecting domains soon after registration. %

In this work we try to quantify, using
public and private sources of data, the gaps in the current models of
data sharing, both regulatory and market-driven.
We focus on domains that are created and removed
in under 24 hours, and thus do not appear in the CZDS zone files.
The likely predominant reason for such \emph{transient} domains is a fraudulent
or malicious activity.
For example, if registrars receive evidence of
malicious use or flag potentially fraudulent payment methods after
they process the registration, the registrar will remove the
domain from the zone immediately \ie before the next CZDS
zone snapshot.
Unfortunately, abuse prevention currently relies solely on the registrars. Each
registrar must independently identify the same signals to block the same threat
actors, who continuously move across registrars to evade detection. This results
in an endless cat-and-mouse game where attackers maintain the upper hand.
Fine-grained visibility into TLD zone changes can enable verified third
parties such as security researchers to build signals to prevent abuse and raise
the cost for threat actors.

In using public and private sources of data to get fine 
grained visibility into TLD zone
changes, we can reduce but not completely close the significant
visibility gap. 
Using ground truth from a medium-size European ccTLD registry, we
detected only one-third of these transient domains
even with the best public available data.
Given the measurable expansion of cybercrime, and the
failure of current data sharing models to effectively combat it,
security-conscious TLDs should consider offering a subscription
service to a feed of changes to zones immediately as they occur.
Verisign offered such a service for the .com TLD
15 years ago \cite{vrsn-rzu-2007} but later terminated it due to its
potential for abuse.  We recognize these concerns of abuse but
believe now is the time to develop a framework to manage these concerns,
just as ICANN is doing for access to even more sensitive
types of registration data \cite{icann-rdrs}.

To that end, we make the following contributions:
\begin{enumerate}
\item We develop, apply, and validate a method that uses CT logs
to estimate a lower
bound on the visibility gap in the daily CZDS snapshots, 
a blind spot for defenders that even commercial
threat intelligence sources do not capture.
\item We provide an open data feed of newly registered domains
(including transient) we discover~\cite{zonestream}. 
\item We investigate the infrastructural landscape of transient domains and the
possible reasons for their early removal.
\item We discuss how to better support transparency
while balancing privacy and commercial interests.
\end{enumerate}

\section{Related Work}
\label{sec:related}

Defenders of networked infrastructure have long had incentives to 
improve detection of malicious domains early in their lifecycle.
A 2020 study found that the average phishing campaign from
start to the last victim took 21 hours~\cite{oest-phishlifecycle-usenixsec20}. 
A parallel study found phishing indicators that blocklists
were unable to detect, concluding that better protection
against phishing would require expansion of
evidence-based reporting protocols \cite{oest-phishtime-usenixsec20}.

Barron \etal~\cite{barronraid2019} found significant correlation
between early domain name deletions and potentially malicious
activities.
Other studies found that spammer domains generally had 
short lifetimes, and that blocklists generally did not detect
them until they were actively used and reported rather than
when initially registered~\cite{hao-imc13}.
Affinito \etal~\cite{domain-lifetime-akiwate-tma-22} found
patterns of suspiciously short-lived domain names in some
TLDs (\eg 80\% under \dns{.xyz}), many of which were reported as
malicious.  Their study revealed that domains registered under
some TLDs (\eg \dns{.xyz}, \dns{.icu}) are taken down within a few days,
while others (e.g., \dns{.top}) remain in the zone longer.

Commercial threat intelligence services have tried to narrow
this visibility gap by monitoring DNS queries in traffic.
In 2018 Domain Tools reported that most newly observed domains
become inactive after 4 hours~\cite{domain-mortality-2018},
presumably because they have been detected and flagged as
malicious and no longer serve their purpose.  Observing DNS
queries from their CDN, Akamai identified 20.1\% (13M)
of all newly registered domain they detected as
malicious.
\cite{akamai}.  In 2022, Palo Alto Newtorks reported that
70\% of newly registered domains detected from passive DNS
data and zone files across 1530 TLDs were
\textit{malicious} or \textit{suspicious} or \textit{not safe for work}~\cite{paloalto}.

In the spike of COVID-related fraudulent online activity, a
group of threat intelligence providers temporarily
volunteered to share threat data in hopes of achieving better
coverage. They found that for novel abuse types,
the aggregated threat intelligence
detected signals that registrars and registries did not~\cite{bouwman-usenixsec-22}.

A recent study of Sommese \etal ~\cite{sommese2023local} 
is relevant to our goal of detection of newly
observed domains without access to proprietary data.
They demonstrated that CT logs are
a reliable and up-to-date source of domain names for addressing
the visibility gaps in ccTLDs, most of which do not participate
in CZDS. They used CT logs to independently
reconstruct more than half of the domains in several ccTLDs
for which they had the full zone file as ground truth.  Their work focused
on spatial coverage whereas we use CT logs to
analyze temporal gaps, i.e., how quickly we detect new and
transient domains.

\section{Methodology}
\label{sec:methodology}

Our approach to estimating the extent of newly registered domains not
captured by daily zone snapshots relies on public data
sources: available zone snapshots, certificate transparency logs~\cite{rfc6962},
active DNS resolution measurements, and registration data
(RDAP)~\cite{rfc7482}.
Unlike passive DNS, our approach does not require the 
domain to be actively queried by users, but the
the domain must have an issued certificate. Our pipeline has
five steps: identify registered (pay-level) domains for which certificates are issued
but are not present in the latest CZDS snapshots; collect RDAP data;
monitor DNS changes of newly observed domains by performing
active measurements; cross-validate inference with RDAP data;
and identify transient (short-lived) domains.
We feed the results of each measurement into Kafka topics
and store them in Parquet format in our object storage
for longitudinal analysis.

\paragraph{Step 1: Infer newly registered domains from 
Certificate Transparency logs.}
We start with the daily zone files extracted by our collector,
which is populated with all latest zones snapshots available from ICANN CZDS.
We use the open-source Certstream \cite{certstream} to
captures logs of newly issued certificates,~\footnote{
We only consider PreCertificate entries, because they 
must always be published before the issuance of the final 
certificate~\cite{rfc6962}} and extract domain 
names from the Common Name (CN) and the Subject Alternative Name 
(SAN) fields of these certificates.  We discard domain names 
that already appeared in the latest daily zone snapshots. This step has three
limitations. First, some CAs do not always perform Domain 
Validation step\footnote{Per Section 4.2.1 of the CA Browser
Forum baseline requirements~\cite{cabrowser}, a CA may reuse
cached validation information collected no more than
398 days prior to issuing the Certificate.}~\cite{ma-staletls-imc23},
so domains with certificates fewer than 398 days old may
no longer exist (\S\ref{sec:results}).
Second, zone file publication may be delayed by days, 
leading to inaccurate inference of domain existence.
Lastly, we are able to detect domains only if a related
certificate is issued.
This step yields a stream of potentially newly registered domain 
names that we feed into our analysis pipeline.
While, this stream contains only domains with an associated issued certificate, 
it represents a publicly accessible data feed for researchers, in contrast with
commercial passive DNS feeds. 
We release this feed publicly at \cite{zonestream}.

\paragraph{Step 2: Collect RDAP registration data.}
We verify how accurately we detect a newly registered domain using
its RDAP-reported registration date and time. In addition, we collect
registrar and registrant information, the latter of which is 
typically redacted to protect privacy.
To collect RDAP data, we
deploy a script in Azure cloud that retrieves newly registered domains
from our Kafka topic and uses the \texttt{whoisit} Python library to
perform RDAP queries for each domain. We leverage Azure functions to
avoid aggressive rate-limiting by cycling measurements over different 
IPv4 addresses. To minimize overhead, we did not retry failed queries.

\paragraph{Step 3: Monitor changes to hosting and DNS providers 
for newly observed domains} 
A reactive measurement infrastructure powered by 
our Kafka streaming pipeline issues DNS 
(A, AAAA, and NS) queries every 10 minutes to these domains
for the first 48 hours of their existence. 
Sixteen instances of our measurement worker,
backed by Unbound (a caching resolver),
execute these queries from 16 separate
nodes.  Our resolver is configured with a
maximum 60-second cache to avoid cached A and AAAA records.
For NS queries, the measurement workers send queries directly to
the domain's TLD authoritative nameserver to more accurately infer
domain removal from the zone, and to prevent misclassification of 
lame delegated or misconfigured domain names as deleted.

\paragraph{Step 4: Validate inference against RDAP data.}
We check that the RDAP registration timestamp is consistent
(within 24 hours) with the observed certificate timestamp.
The difference between these two timestamps 
provides a ground truth indication of how correctly and 
promptly we detect newly registered domain names.   
This step only works when our RDAP resolution 
succeeded ($\approx 97\%$ of cases\footnote{We missed
$\approx3\%$ due to RDAP rate limiting or other collection errors.}).

\paragraph{Step 5: Identify transient (short-lived) domains,}
which we define as those registered and deleted in the time
span (max 24 hours) between the two zone snapshots.  
Note that we do not include short-lifetime domains that the
zone snapshots do manage to capture.  In our data
set, these transient domains are the subset of newly registered
domains that do not appear in any zone file throughout our
analysis window (1 Nov 2023 - 31 Jan 2024).

\section{Results}
\label{sec:results}

\begin{table*}[h]
\centering
\caption{Top 10 TLDs ranked by count of newly registered domains (NRDs) for Nov 23 - Jan 24).  The Zone NRD columns shows how many NRDs appeared in the daily zone file snapshots.}
\begin{tabular}{lrrrrrr}
\toprule
TLD Category & Nov & Dec & Jan & Total & Zone NRD & Coverage NRD (\%) \\
\midrule
com & \num{1127727} & \num{1109804} & \num{1505044} & \num{3742575} & \num{8467641} & 44.2\% \\
xyz & \num{114582} & \num{87051} & \num{107740} & \num{309373} & \num{649010} & 47.7\% \\
shop & \num{76626} & \num{99660} & \num{107675} & \num{283961} & \num{775253} & 36.6\% \\
online & \num{76674} & \num{76693} & \num{109964} & \num{263331} & \num{648922} & 40.6\% \\
bond & \num{75779} & \num{81265} & \num{84997} & \num{242041} & \num{292552} & 82.7\% \\
top & \num{82746} & \num{74134} & \num{83837} & \num{240717} & \num{532363} & 45.2\% \\
net & \num{79660} & \num{71922} & \num{84320} & \num{235902} & \num{643030} & 36.7\% \\
org & \num{53377} & \num{53767} & \num{76400} & \num{183544} & \num{481870} & 38.1\% \\
site & \num{46695} & \num{47879} & \num{65801} & \num{160375} & \num{465542} & 34.4\% \\
store & \num{42931} & \num{38699} & \num{50279} & \num{131909} & \num{326383} & 40.4\% \\
Others & \num{328570} & \num{333000} & \num{380551} & \num{1042121} & \num{3009575} & 34.6\% \\
Total & \num{2105367} & \num{2073874} & \num{2656608} & \num{6835849} & \num{16292141} & 42.0\% \\
\bottomrule
\end{tabular}
\label{tab:agg_stat_nrd}
\end{table*}

Recall, our methodology uses CT logs to identify \emph{use of domains
before} they show up in the CZDS zone snapshots. From 1 Nov 2023 to 31 Jan
2024, we found 6.8 million domain names (\autoref{tab:agg_stat_nrd}) that
did not appear in the CZDS zone snapshot before we saw them in the
CT logs. These $6.8M$ domain names 
are either transient domains or eventually show up in the next CZDS zone
snapshots. Most of these domains (3.7M) were in \dns{.com}, followed 
by \dns{.xyz} (300K) and \dns{.shop} (284K).

When comparing our list of detected newly registered domains to
the diff between two daily CZDS zone snapshots, we found that our methodology
identified 42\% of the newly registered domains before they
were published in the zone snapshots in our 3-month observation period.
The remaining 58\% did not have certificates issued immediately.
Since our goal is to identify transient domains \ie domains that
do not show up in zone snapshots, this comparison is purely functional.

\subsection{Detection Speed}
\begin{figure}[t]
    \centering
    \includegraphics[width=0.9\columnwidth]{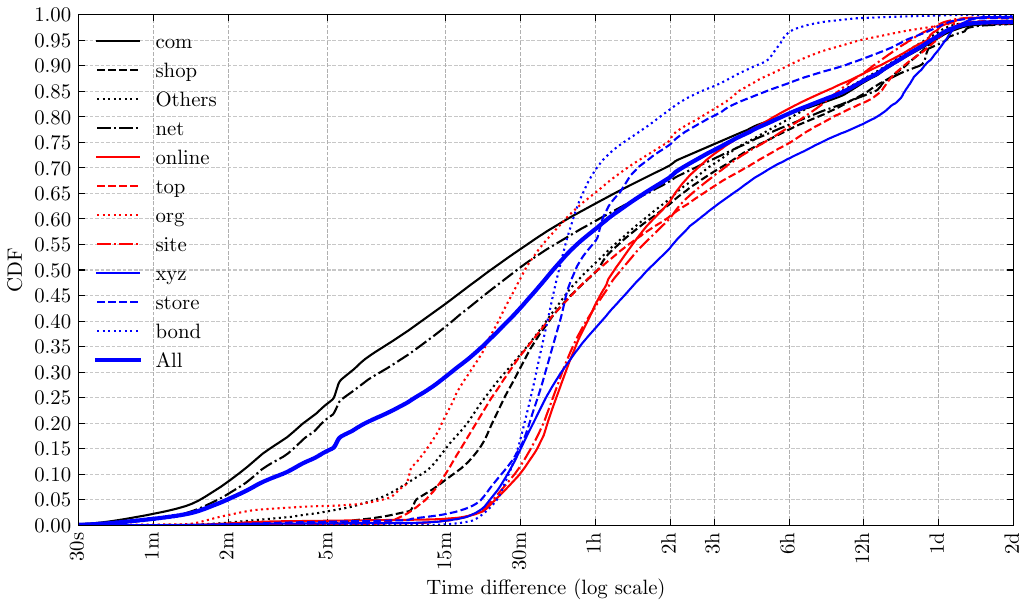}
    \caption{Difference in registration time per RDAP vs.~CT logs.
	We detected 50\% of domains within 45 minutes of their
	existence, and $\approx$30\% within 15 min.}
    \label{fig:lag}
\end{figure}

\autoref{fig:lag} plots the time difference between 
when we recorded the domain (\ie fetched it from
CT Logs\footnote{We use the Certstream-reported timestamp because
neither precertificates nor CT logs provide an insert timestamp})
and the RDAP-reported timestamp. 
We detected half of the newly registered 
domains within the first 45 minutes of their
existence, and $\approx$30\% within the first 15 minutes.
The distribution has a long tail toward 1 day
(less than 2\% have a difference greater than a day), 
which we believe derives from our misclassification of newly
registered domains (\eg due to incorrect extraction of the
second-level domain (SLD) using the Public Suffix List) or zones published with several
days of delay.
The time differences across TLDs can be explained
by the operational update times of their zones (\ie the domain 
needs to be active in the zone before a certificate can be issued).
For example, the \dns{.com} and \dns{.net} registries update these
TLD zones on average every 
60 seconds while other gTLDs registries update their
zones every 15-30 minutes.
We validated this assumption
by probing the zones of \autoref{fig:lag}
for SOA serial changes, and found consistent timestamps.

This result demonstrates that our methodology correctly and
quickly identifies newly registered domains. The RDAP data 
helps to exclude corner cases and misclassified (as newly registered)
domain names, and provides relevant metadata (\eg registrar
identity, accurate creation time).

Furthermore, our reactive measurements allowed us to provide 
a lower-bound estimation of DNS infrastructural changes. 
In our data set, most (97.5\% of) newly 
registered domains kept their initial nameserver infrastructure
for the first 24 hours.  A few (2.5\%) changed NS infrastructure
within the first 24 hours such that the daily zone snapshot differences 
could miss it, depending when in the 24-hour window the change occurred.

\subsection{Transient Domains} While early detection of newly
registered domains represents a significant advance 
detecting possible malicious behavior, those domains will generally 
show up in
the zone snapshots in the days following their registration.
However, a small percentage (1\%) of newly registered domains
never appeared in a zone snapshot due to their short lifespan
falling between two snapshots. We call these domains {\em
transient domains}.  We identified these names by excluding
from our list of newly registered domains those that appeared
in our zone collection during the 3 month observation window
(+/- 3 days to account for late zone snapshots publishing).

\begin{table}[h] 
\centering 
\caption{Transient domain names observed} 
\begin{tabular}{lrrrr} 
\toprule TLD Category & Nov & Dec & Jan & Total \\ 
\midrule com & \num{9363} & \num{10597} & \num{21232} & \num{41192} \\ 
online & \num{1800} & \num{2369} & \num{1990} & \num{6159} \\ 
site & \num{1578} & \num{1381} & \num{890} & \num{3849} \\ 
net & \num{702} & \num{866} & \num{1544} & \num{3112} \\ 
org & \num{595} & \num{602} & \num{1176} & \num{2373} \\ 
shop & \num{688} & \num{497} & \num{507} & \num{1692} \\ 
xyz & \num{321} & \num{316} & \num{624} & \num{1261} \\ 
store & \num{422} & \num{414} & \num{377} & \num{1213} \\ 
top & \num{213} & \num{161} & \num{276} & \num{650} \\ 
fun & \num{185} & \num{175} & \num{160} & \num{520} \\ 
Others & \num{1609} & \num{1958} & \num{2454} & \num{6021} \\ 
Total & \num{17476} & \num{19336} & \num{31230} & \num{68042} \\ 
\bottomrule \end{tabular} 
\label{tab:agg_stat_transient} 
\end{table}

Our method inferred a lower bound of 68,042 transient domains
(\autoref{tab:agg_stat_transient}), i.e.,
$\approx$1\% of CT-observed newly registered 
domain names. 
The TLD share of these two populations (transient and 
newly registered) differs (\autoref{tab:agg_stat_nrd}
vs.~\autoref{tab:agg_stat_transient}), perhaps
related to TLD pricing oscillation and bulk 
malicious registration campaigns 
\cite{interisle-2023,cybercrimesunrise}.

During our data collection, we noticed that
RDAP failure rate of transient domains was noticeably higher
($\approx$34\%)
than for newly observed domains ($\approx$3\%). 
We identified three major causes for this
higher failure rate: (i) we detected too late, i.e., when we attempted to collect
the RDAP data, the domain had already been removed; (ii)
we were too early, \ie RDAP data was not yet in sync, and (iii)
we detected domains that no longer existed but for which a
certificate had been issued.

While we cannot investigate causes (i) and (ii) due to lack
of RDAP data, we examined several
domains in the third category, contacting the CERT teams of
the CAs that issued certificates for these non-existent domains.
GlobalSign, Sectigo, and Cloudflare confirmed that issuing a
certificate for a non-existent domain (more precisely,
issuing a certificate without validating the existence of the
domain) is allowed under the condition that the CA possesses
a previously obtained and valid Domain Validation (DV) token
(see \S\ref{sec:methodology} and footnote 2).
A necessary but not sufficient condition
for this case is that the domain existed in the past.  We
conducted a comparison of the 34\% of transient
domains for which we failed to collect RDAP data
with the CAIDA DZDB historical zone collection~\cite{dzdb} and found that
approximately 97\% of those domains were registered in the
past. 
Another possible approach to filter out these domains
is to individually check if a certificate was issued in the
DV token validity period in the past.   
Given the complexity of longitudinal 
investigation of CT Logs, we leave this filtering as future work.

Since domains with non-responding RDAP may not be transient (cause iii),
we filtered them from subsequent analysis. 
Using the RDAP-reported registration timestamp, 
we also filtered domains misclassified as newly registered.
This filtering yielded 42358 confirmed transient domains over the 
3-month period.
These transient domains represent only a portion of all the transient
domains across the zones we analyzed. Because we do not have access to
ground truth (\ie registries view), we cannot establish the total number
of transient domains or the extent of our methodology's coverage. 
In \autoref{sec:visgap},  we will compare our detection with a passive 
DNS source and the ground truth from a ccTLD.

\subsubsection{Lifetimes of Transient Domains}

\begin{figure}[t]
    \centering
    \includegraphics[width=0.85\columnwidth]{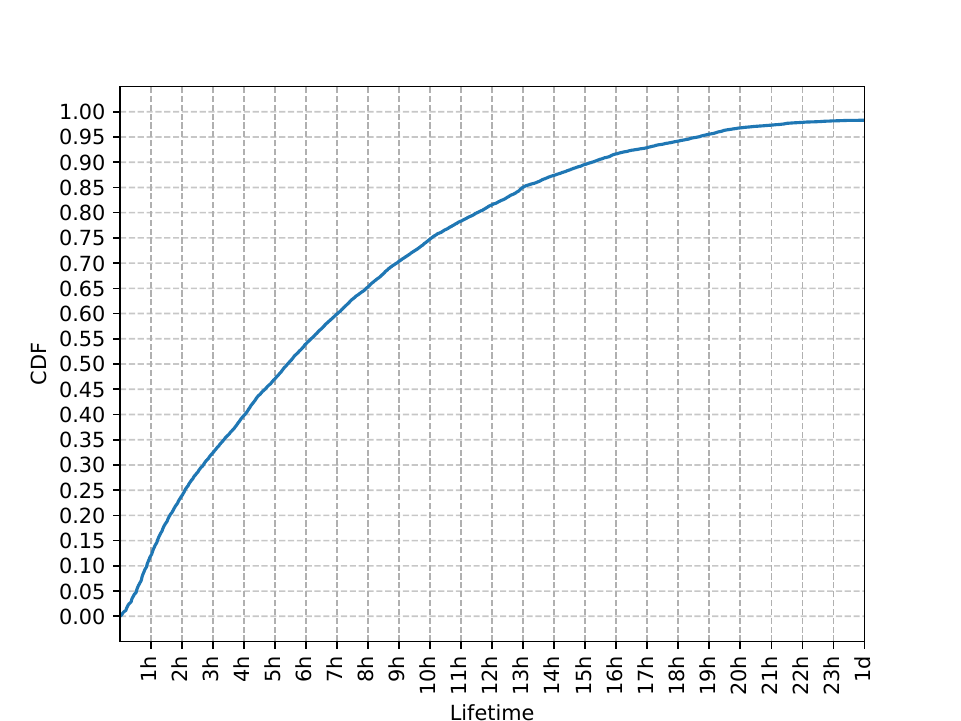}
    \caption{Lifetime of transient domain names.}
    \label{fig:lifetime}
\end{figure}

To analyze how quickly transient domains disappear from
their zone, we subtracted the last time the TLD
nameserver provided a valid response for the NS query for that
specific domain from the RDAP registration time of the domain.
Per this method, over 50\% of these domain names died within
their first 6 hours of life (\autoref{fig:lifetime}).

\subsubsection{Registrar and Hosting Landscape of Transient Domains}

\begin{table}[t]
\centering
\caption{Top 10 Transient Domains Registrars Distribution.}
\begin{tabular}{lcc}
\toprule
Registrar & Domains & \% \\
\midrule
GoDaddy & 8213 & 19.39\% \\
Hostinger & 6418 & 15.2\% \\
NameCheap & 4195 & 9.9\% \\
Squarespace & 2820 & 6.7\% \\
Public Domain Registry & 2625 & 6.2\% \\
IONOS & 2352 & 5.6\% \\
Metaregistrar & 1866 & 4.4\% \\
NameSilo & 1853 & 4.4\% \\
Network Solutions, LLC & 1670 & 3.9\% \\
Tucows & 1304 & 3.1\% \\
Others & 9042 & 21.3\% \\
\midrule
Total & 42358 & - \\
\bottomrule
\end{tabular}
\label{tab:registrar_transient}
\end{table}

The collected RDAP data revealed that market leader 
GoDaddy topped the list of registrars holding transient 
domains, with 19\% of such domains
(\autoref{tab:registrar_transient}).
The registrar landscape is dominated by large registrars, with
smaller registrars making up the long tail of the distribution
(see Others in \autoref{tab:registrar_transient}).  These 
results suggest that transient domains are a widespread
phenomenon across registrars.
\begin{table}[b]
\centering
\caption{Top 5 DNS Hosting Domains (NS records) of Transient Domains}
\begin{tabular}{llrr}
\toprule
Name & NS Record SLD & Domains & \% \\
\midrule
Cloudflare & cloudflare.com  & 20981 & 49.5\% \\ 
Hostinger & dns-parking.com & 3682 & 8.7\% \\ 
NS1 & nsone.net & 2938 & 6,9\% \\
Squarespace & squarespacedns.com & 2908 & 6.9\% \\
GoDaddy & domaincontrol.com & 2315 & 5.5\% \\ 
Others & - & 9534 & 22.5\% \\
\midrule
Total & - & 42358 & - \\
\bottomrule
\end{tabular}
\label{tab:dnshosting_transient}
\end{table}

\begin{table}[t]
\centering
\caption{Top 5 Web Hosting (A records) of Transient Domains}
\begin{tabular}{llrr}
\toprule
Name & ASN & Domains & \% \\ 
\midrule
Cloudflare & 13335 & 15322 & 36.2\% \\ 
Hostinger & 47583 & 5930 & 14.0\% \\ 
Amazon & 16509 & 3198 & 7.6\% \\ 
Squarespace & 53831 & 2257 & 5.3\% \\ 
Namecheap & 22612 & 1650 & 3.9\% \\ 
Others & - & 14001 & 33.1\% \\ 
\midrule
Total & - & 42358 & - \\
\bottomrule
\end{tabular}
\label{tab:hosting_transient}
\end{table}

We used our active resolution measurements to analyze the 
DNS hosting (\autoref{tab:dnshosting_transient}) and web hosting 
(\autoref{tab:hosting_transient}) infrastructure of transient domains. 
For DNS hosting, we investigated the most popular nameservers SLDs, while for
Web hosting we looked at the ASNs of the A records of the domains.
Half of these transient domains were using Cloudflare as DNS provider (\ie
for their authoritative nameservers) and $\approx$35\% of them
used Cloudflare as a CDN provider. 
We identified Hostinger's parking DNS nameservers as the second
most prominent DNS provider, as accounting for $\approx$8\% of the transient
domains, followed by a long tail of popular DNS hosting services
including NS1, Squarespace, and GoDaddy.  
These findings suggest that those domains are generally hosted on
similar infrastructure of long-lived domains.  In contrast, the large
presence of parked domains could indicate domains that have already been
removed or caught before misuse.

\subsection{Reasons for Early Removals}

In private conversations with two top registrars in
\autoref{tab:registrar_transient}, they confirmed that
transient domains are mostly likely malicious.
The registrars did not comment on exact numbers but anecdotally acknowledged 
that with few exceptions, reasons for early removal include abuse, account suspensions,
or credit card fraud. Some legitimate cases are domain tasting and right of cancellation,
but they are exceptionally rare.
To investigate indicators of maliciousness of transient domain names, 
we analyzed how often blocklists manage to flag them.
We considered both early-removed newly registered domain names
and 42358 transient domains (\autoref{tab:registrar_transient}).
With \textit{early-removed}, we refer to newly registered domains that 
were removed before the end of our analysis period.
Unlike transient domains, early-removed domains appear in the zone file
snapshots but still have a shorter lifespan than the 1-year typical duration of
a registered domain due to the likely forced removal by the registrar~\cite{domain-lifetime-akiwate-tma-22}.
We observed 555491 domains that were deleted before
1 February 2024 (10\% of newly registered
domains we detected).

We examined ten blocklists daily from 1 November 2023, to 29 April 
2024, extending beyond our observation period to 
capture late insertions: DBL~\cite{dbl},
Phishtank~\cite{phishtank}, Phishingarmy~\cite{phishingarmy},
Cybercrimetracker~\cite{cybercrimetracker}, Tolouse {DDoS,
Crypto, Malware}~\cite{tolouse}, Digitalside~\cite{digitalside},
Openphish~\cite{openphish}, Vxvault~\cite{vxvault},
Ponmocup~\cite{ponmocup}, Quidsup~\cite{quidsup}.

\paragraph{Newly registered domains.}
Of 555491 newly registered domains, 
at least one blocklist in our set identified 
6.6\% (37188) as malicious. Of these, 92\% (34233) were active
when we detected them, 3\% (1072) were already included
in the blocklists before their registration date, and 5\%
(1882) appeared on blocklists after deletion.

\paragraph{Transient domains.} Of the 42358
transient domains, at least one blocklist flagged 5\% (2123) 
as malicious. Of these, 5\% (105) were flagged on their 
registration date, 1\% (12) were already on the blocklists
before their registration date, and 94\% (2006) appeared 
after domain deletion.

\paragraph{Summary} 
Blocklists identify and label as malicious a tiny percentage
of domains before their registration dates, which may indicate
a re-registration.  Additionally, a significant percentage
(94\%) of transient malicious domains appeared on the blocklists
post-deletion, indicating that blocklists do not promptly or
in some cases ever detect short-lived malicious domains
\cite{bouwman-usenixsec-22}.

\subsection{Visibility Gap in Domain Detection}
\label{sec:visgap}

Commercial threat intelligence services~\cite{dt-sie} using passive DNS
have tried to fill the gap by detecting domains soon after registration,
and are generally the best available data source for identifying
transient domains. To understand the differences between identified 
transient
domains using public and private data sources, we collected and compared one day (May 9,
2024) worth of newly registered domains provided by DomainTools's SIE Newly
Observed Domain (NOD) feed~\cite{dt-sie} with our feed.
To simplify the comparison and avoid
accounting for domains detected after several days, we compared only domains
that were registered on May 9, 2024, according to the RDAP data from both our
feed and the SIE feed.  We considered only gTLD domains because they were
available in both data feeds.

We first compared newly registered domains, finding that the 
SIE NOD feed detected $\approx$5\% more domains than our 
method. However, the overlap between the two data sources is 
$\approx$60\%, indicating that each method detects a unique
subset of domains.
Looking at transient domains identified by the two
sources, we find the overlap drops further. A total of 855 transient
domains 
were identified by either one of the two sources. Of these, only 33\% were
detected by both, with the SIE NOD feed identifying 10\% more transient
than our method. While the SIE NOD detects a larger number,
the disjoint nature of the intersecting set of domains suggests the need
to combine them to narrow the visibility gap.

To validate our visibility gap
against a source of ground truth, we use the perspective 
of a mid-sized ccTLD registry (\texttt{.nl}) on transient domains. 
From November 2023 until
January 2024, the registry observed 714 domain names that were
deleted in less than 24 hours in their registration system.
Of those domains, 334 were registered and deleted such that
they were never captured in zone file snapshots.  
Applying our methodology to this ccTLD, we found only 99 transient 
domains, or 29.6\% of the 334 
ccTLD-identified transient domain names.\footnote{We were unable to
compare the SIE NOD feed to the registry data because we had 
access only to one day's worth of data.
However, based on our previous results we argue that combining the two data sources
would still leave a large visibility gap.}
This result shows that while we can catch
a fraction of these transient domains,
researchers still have a huge blindspot regarding
intra-day events.

\section{Discussion and Conclusion}

It is well-established that domain-related cybercrimes generally run their
course or affect the most victims within 24 hours of attack onset
\cite{cybercrime-interisle-2023,hao-imc13,oest-phishlifecycle-usenixsec20,domain-mortality-2018}.
The phenomenon of transient domains indicates some measure of
success that registrars detect and take down so many short-lived
domains in their early stages, before they can do damage.
However, in this current model, each registrar has to independently
re-learn the same signals as threat actors move across different
registrars to evade detection. 
This state of affairs not only inflicts additional cost on registrars but also
increases the time to remediate abuse. 
Even well-intended registrars and registries cannot always
investigate reports of abuse in a timely manner, and the process
for doing so is not well-defined \cite{cybercrime-interisle-2023}.

In the meantime, transient domains, in which malicious activity
tends to dominate, have largely been invisible to the research community.
At the same time, research has established that the existing DNS-related threat 
intelligence community, \eg blocklist companies, are less effective 
against new and emerging threats \cite{bouwman-usenixsec-22},
suggesting the need for new approaches.

\noindent
\paragraph{Learning from history}
Verisign long ago supported
{\em rapid zone updates}, enabling updates to the .com and
.net zone every 5 minutes.  The data included
domain names, nameservers, IP address additions, deletions and
modifications, to ``promote security and stability by providing
a useful tool to online security companies, ISPs, search
engines, financial services companies, and other stakeholders''
~\cite{vrsn-rzu-2007},~[\ref{appendix:rzu}].
This service illuminated the activity of bad actors,
including phishing, fraud and identity theft \cite{hao-imc13}.
Concerns about abuse of the data by spammers led to Verisign's
termination of this service.   In particular, attackers exploited
a brief time window where the zone updates were publicly available, 
but before domain owners had set up protection of their domain.
\paragraph{Resurrecting RZU} 
Given the inherent advantages that attackers have over defenders, 
the ineffectiveness of existing uncoordinated countermeasures,
and the limited obligations (or ability) of registrars
to mitigate harm,\footnote{Interisle reports that ICANN's recently 
revised contract language does not attribute any duty of care to 
registrars or registries, and does not address vulnerabilities such 
as exact-match brand registration and suspicious bulk registration
\cite{cybercrime-interisle-2023}.}
we see a need to expand transparency and
accountability mechanisms. 

Registrars and registries who want to establish themselves as serious about
security could resurrect the capability to subscribe to rapid zone updates along
with a framework to safeguard against abuses, and to learn from history how to
mitigate the risk of abuse. 
While risks of abuse of access to non-public DNS data has created tension and
controversy for decades, the CZDS program is testament to the ability to manage
these risks.  More recently, ICANN has coordinated a model 
for requesting access to other non-public registration data 
in a more consistent and standardized format 
\cite{icann-rdrs}.
We advocate for the development of a similar model for data sharing among approved 
and trusted parties, including operators, law enforcement, and the research community.
Such access to rapid zone updates
would enable a source of labeled data that would
allow application of machine learning techniques not only to
security research on systemic harms, but also to anti-abuse
efforts of registries, registrars, and law enforcement agencies.

\paragraph{Future Directions} 
To demonstrate the visibility gap of daily zone snapshots,
we proposed a methodology to identify newly registered domains 
based on CT logs and active measurements, which we released
as a public stream~\cite{zonestream}. Our results shed light
on previously undetected short-lived domains, often related
to malicious activities, which we named transient domains.
Preliminary analysis and contact with registrars suggest these
domains are likely malicious.   In the future we plan to 
expand our measurements beyond DNS infrastructure
records, including mail extensions (\eg \texttt{SPF}, \texttt{MX},
\etc), subdomains, and web-crawling.  These measurements,
combined with early detection using our methodolgy, can
support machine learning-based approaches to proactively
identify malicious domains before they do harm. 
Proactive registries can leverage such use cases to 
justify and sustain a responsible RZU service.

\bibliographystyle{ACM-Reference-Format}
\balance
\bibliography{bib}
\appendix
\section*{Acknowledgments}
We thank our anonymous shepherd and IMC reviewers for their
insightful suggestions and feedback.
We thank Cristian Hesselman and Roland van Rijswijk-Deij for their support and insights over time.
Additionally, we extend our thanks to Domain Tools and SIDN Labs for providing us with access to 
passive DNS feeds and registry ground truth.
This material is based on research sponsored by the National Science Foundation (NSF) grant OAC-2131987.
The views and conclusions contained herein are those of the authors and should not be interpreted as 
necessarily representing the official policies or endorsements, either expressed or implied, of NSF.
This work was partially supported by the G\'EANT GN5-1 programme funded by the European Commission.
This research was made possible by OpenINTEL~\cite{oijsac}, a joint project of the University of Twente,
SURF, SIDN, and NLnet Labs; and operational support from the Twente University Centre for
Cybersecurity Research (TUCCR).

\section{Ethics}
We acknowledge that making a public feed of newly registered domain
names available may expose personal information (\eg personal name 
used as a domain name). However, we argue that our
method only consolidates information already publicly
available in the CT Logs ecosystem. Therefore, we decided
not to  remove any domain names from the published stream.

Our ethics review board does not require a specific ethics approval
for internet measurements conducted according to community best practices. 
In measuring the DNS infrastructure of newly registered
domains, we followed these best practices by aiming
to strike a balance between coverage and slower probing rate.
As a result, we decided to scan those domains every 10 minutes, with only three
infrastructural query type (\texttt{A}, \texttt{AAAA}, \texttt{NS}), and without enumerating possible sub-labels. 
Furthermore, we ensured that the address space of our
measurement infrastructure had pointers to direct
interested users to a page explaining the project's
goals and providing a contact for opting out from the
measurements.

Regarding RDAP measurements, we queried approximately 76,000 domains per day. 
We limited our measurements to a maximum of one domain per second 
to accommodate possible bursts. This rate is generally acceptable and 
falls well below most registry rate limits (\eg 7,200 queries per hour 
for CentralNic) for RDAP. We distributed measurements across four workers
with different IPs, to avoid rare cases of registry RDAP servers enforcing
extremely aggressive rate limiting and permanently blocking IPs.
To prevent overburdening the infrastructure, we purposefully avoided
repeating RDAP measurements in case of failure. We monitored
the failure rate to ensure that we did not overload the targeted servers.

\section{Appendix: Verisign's request to provide Rapid Zone updates}
\label{appendix:rzu}

The following is a quotation from Verisign's application to
ICANN to provide rapid zone updates as a service
~\cite{vrsn-rzu-2007}:

In September 2004, VeriSign implemented rapid zone updates, enabling
updates to the .COM and .NET zone every 3 minutes (prior to this VeriSign
propagated updates to the .COM and .NET zones every 12 hours). Although
VeriSign updates these zones every 3 minutes, VeriSign publishes that
updated data twice a day.  This data includes domain names, nameservers,
IP address additions, deletions and modifications. The proposed service
would enable registrars and others (i.e., anyone who wishes) who currently
obtain zone file access in the .COM and .NET TLDs twice daily, to receive
updated zone information every five minutes.

....

VeriSign states that the service would be used by recipients to build
brand protection and fraud detection services for their customers, and
promote security and stability by providing a useful tool to online
security companies, ISPs, search engines, financial services companies,
and other stakeholders.

\begin{itemize}
\item The service would shed light on the activity of those
engaged in domain tasting and expose bad actors. ``This [service]
elevates bad actors into the light of day.'' ``All it does is
out people who are probably bad guys.''

\item The service does not seem susceptible to gaming;
``it prevents gaming.''

\item The service will increase choice for registrars: they can continue
to obtain information in twelve-hour increments, purchase the service
directly from VeriSign, or purchase the service from value-added providers.

\item The service would provide ``more granularity'' regarding tasting
activities but not impact the practice significantly.

\item Phishers and others currently alter name servers and conduct
fraudulent activity during the time between the twice daily publication
of zone file updates.  That is, they make nameserver changes to conduct
fraudulent activity immediately after one publication and make the change
back prior to the next publication -- there is a twelve-hour window to
conduct this activity.

\item Intellectual property owners, brand protection managers and law
enforcement would be able to improve their ability to search the .COM
and .NET zones for ``typo-squatters''. ``The service would advantage law
enforcement to thwart phishing.'' (One opinion was that the additional
twelve-hour extra notice period might not be a significant advantage.)

\item After consideration and attempts to, ``poke holes in it,''
there was no apparent ability to use the service for ``bad''
purposes.
\end{itemize}

Comments from other sectors of the community have indicated that VeriSign's
proposal would assist those who use zone file information provided by
VeriSign to address such problems as phishing, fraud and identity theft.

External counsel and ICANN staff agreed that there are no apparent
competition issues that require that this matter be forwarded to relevant
competition authorities at this time.

\end{document}